\documentclass[aps,pra,twocolumn,amsfonts,amssymb,amsmath,showpacs,
floatfix,nofootinbib,groupedaddress,superscriptaddress,citesort]{revtex4}
\usepackage{mathrsfs}
\usepackage{amsfonts}
\usepackage{amstext}
\usepackage{amsmath}
\usepackage{amssymb}
\usepackage[dvips]{graphicx}
\def\qed{\leavevmode\unskip\penalty9999 \hbox{}\nobreak\hfill
     \quad\hbox{\leavevmode  \hbox to.77778em{%
              \hfil\vrule   \vbox to.675em%
               {\hrule width.6em\vfil\hrule}\vrule\hfil}}
     \par\vskip3pt}

\newtheorem{theorem}{Theorem}

\newtheorem{cor}[theorem]{Corollary}

\begin{document}
\title{Evolution equation for entanglement of assistance}

\author{Zong-Guo Li}
\affiliation{Beijing National Laboratory for Condensed Matter
Physics, Institute of Physics, Chinese Academy of Sciences, Beijing
100190, China}
\author{Ming-Jing Zhao}
\affiliation{School of Mathematical Sciences, Capital Normal
University, Beijing 100048, China}
\author{Shao-Ming Fei}
\affiliation{School of Mathematical Sciences, Capital Normal
University, Beijing 100048, China}
\author{W. M. Liu}
\affiliation{Beijing National Laboratory for Condensed Matter
Physics, Institute of Physics, Chinese Academy of Sciences, Beijing
100190, China}

\begin{abstract}
We investigate the time evolution of the entanglement of assistance
when one subsystem undergoes the action of local noisy channels. A
general factorization law is presented  for the evolution equation
of entanglement of assistance. Our results demonstrate that the
dynamics of the entanglement of assistance is determined by the
action of a noisy channel on the pure maximally entangled state, in
which the entanglement reduction turns out to be universal for all
quantum states entering the channel. This single quantity will make
it easy to characterize the entanglement dynamics of entanglement of
assistance under unknown channels in the experimental process of
producing entangled states by assisted entanglement.
\end{abstract}
\maketitle
\section{\bf Introduction}

In quantum information, entanglement is a vital resource for quantum
cryptography and quantum computation \cite{horodecki,nielsen}. More
specially the distribution of bipartite entanglement is a key
ingredient for performing certain quantum-information processing
tasks such as teleportation. Therefore, the creation and
distribution of entanglement is of central interest in quantum
information processing.

There is an alternative to producing bipartite entanglement by reducing a
multipartite entangled state to an entangled state with fewer
parties (e.g. bipartite) via measurements on the other parties. Such producing of
entanglement, called ``assisted entanglement", is a special
case of the \emph{localizable entanglement} \cite{localizable},
which is especially important for quantum communication, where
quantum repeaters are needed to establish bipartite entanglement
over a long length scale \cite{hj1}. The entanglement in this produced
entangled state is quantified by the entanglement of assistance (EOA) which
is defined in Refs. \cite{cohen,dp}.

However, in the transition and store of quantum entangled state, we
inevitably encounter some interactions of the multipartite quantum
state under consideration with its environment. For example, in an
experimental preparation of entanglement for a quantum information
processing task, suppose a supplier of entanglement, named
``Charlie", creates entanglement between two distant parties, Alice
and Bob, by the assisted entanglement. Suppose also that Charlie
starts with a pure tri-qudit entangled state such as the
Greenberger-Horne-Zeilinger (GHZ) state \cite{ghz} and sends, via
quantum channels, two of the three qudits to Alice and Bob. Due to
the environmental interaction, the quantum channels generally make
the pure tripartite entangled state degrade to a mixed tripartite
entangled state. These undesired couplings give rise to decoherence,
which may decrease the entanglement in the created state when the
particles propagate. Therefore, it is of great practical importance
to investigate the dynamics of entanglement of assistance for the
quantum system under the influence of decoherence.

Recently, the entanglement evolution has been studied extensively
for bipartite system under the influence of local decoherence. In
stead of deducing the evolution of entanglement from the time
evolution of the state a direct relationship between the initial and
final entanglement of a bipartite state has been explored in e.g.,
Refs. \cite{thomas,markus,li}. In this paper, we study the time
evolution of EOA when one subsystem undergoes the action of an
arbitrary noisy channel. Our results demonstrate that the dynamics
of EOA is determined by the action of channel on the bipartite pure
maximally entangled state. In particular, for a $2\times 2\times
n_3$ pure quantum state with one qubit subsystem being subject to a
quantum channel, we get a general factorization law for the
evolution equation of EOA, i.e., the EOA of the final state can be
expressed as the product of the initial EOA multiplied by the
entanglement reduction of maximally entangled state after the
process of the quantum channel's action. The latter quantity is
universal for all initial states passing through this channel.
We take generalized amplitude damping and phase
damping channels as examples, and find that the sudden death of EOA
exists in the evolution and is only determined by the entanglement
evolution equation of the maximally entangled state entering this
channel. Moreover, for the other cases with one subsystem undergoing the
action of an arbitrary noisy channel, a similar relation is
satisfied, with a sacrifice that the equality is replaced by an
inequality.

This paper is organized as follows: We start by introducing a
definition of EOA and further study the dynamics of EOA for $2\times
2\times n_3$ systems with a qubit being subject to a noisy channel.
In Sec. III, the evolution equation of EOA is investigated for a
general $d\times d\times n_3$ state with a local operation on the
second subsystem. We apply the evolution equation to some cases and
discuss the result in Sec. IV. Finally, in Sec. V, we conclude with
a summary of our results.

\section{\bf Evolution equation of EOA for $2\times 2\times n_3$ systems}
A 2-qubit entangled state is the basic unit of quantum information.
An important application of assisted entanglement is to create a
2-qubit entangled state from a $2\times 2\times n_3$ entangled
state. Therefore, we firstly consider a pure $2\times 2\times n_3$
state shared by three parties referred to as Alice, Bob and Charlie.
Charlie's aim is to maximize the entanglement of the state between
Alice and Bob by performing a local operation on his system and
communicating the result to Alice and Bob. For $2\otimes2$ systems,
we take concurrence to quantify the entanglement and EOA then reduce
to the concurrence of assistance \cite{laustsen,gour1}. For a pure
$2\times 2\times n_3$ state $|\psi\rangle_{ABC}$, the EOA is defined
as follows:
\begin{eqnarray}
\label{EOA} E_a(|\psi\rangle_{ABC})\!\equiv\!
E_a(\rho_{AB})\!\equiv\!\textrm{max}\sum_ip_iC(|\phi_i\rangle_{AB}),
\end{eqnarray}
where
$\rho_{AB}=\mathrm{Tr}_C\big(|\psi\rangle_{ABC}\langle\psi|\big)$,
and $C(|\phi_i\rangle_{AB})$ is the concurrence defined in Ref.
\cite{wootters}. The maximum runs over all possible pure-state
decompositions of
$\rho_{AB}=\textrm{Tr}_C[|\psi\rangle_{ABC}\langle\psi|]=\sum_ip_i|\phi_i\rangle_{AB}\langle\phi_i|$.
Equation (\ref{EOA}) quantifies the maximum average entanglement
that Charlie can create between Alice and Bob because any pure-state
decompositions of $\rho_{AB}$ can be realized by a generalized
measurement performed by Charlie (for more details see
\cite{gour,hughston}). Thomas Laustsen \emph{et al.} \cite{laustsen}
have given an analytical expression for EOA,
\begin{eqnarray}
\label{EOA0}
E_a\big(|\psi\rangle_{ABC}\big)\!=\!\mathrm{Tr}\sqrt{\sqrt{\rho_{AB}}\tilde{\rho}_{AB}\sqrt{\rho_{AB}}}
\!=\!\sum_i\lambda_i,
\end{eqnarray}
where
$\tilde{\rho}_{AB}=\sigma_y\otimes\sigma_y\rho^*_{AB}\sigma_y\otimes\sigma_y$
with $\sigma_y$ Pauli matrix, and $\lambda_i$ are the square roots
of the eigenvalues of $\rho_{AB}\tilde{\rho}_{AB}$. In a similar
way, for a mixed tripartite state $\rho_{ABC}$, the EOA is defined
as the maximum achievable average concurrence shared by Alice and
Bob after Charlie's assistance, depends on $\rho_{ABC}$ and thus not
solely on $\rho_{AB}$.

Suppose $|\psi\rangle_{ABC}$ be the tripartite pure initial state,
and the second subsystem undergo the action of a noisy channel. We
will denote the noisy channel by $\mathcal{S}$ and take the
positive-operator valued measure (POVM) measurements to describe the
channel $\mathcal{S}$ thereafter. Then the final state of the system
takes the form
$\rho'\!=\!(\mathbf{1}\otimes\mathcal{E}\otimes\mathbf{1})|\psi\rangle_{ABC}\langle\psi|$,
which is usually mixed. In general, the EOA is difficult to solve
for mixed states. For this state $\rho'$ with one qubit having gone
through a noisy channel, we have an analytical expression of EOA as
shown in the following:
\begin{theorem}
For a pure $2\times 2\times n_3$ state $|\psi\rangle_{ABC}$, the second
subsystem of which is undergoing the action of the channel $\mathcal{E}$, the EOC of
the final state $\rho'$ takes the form as the following equation:
\begin{eqnarray}
\label{theo1}
&&E_a\big[(\mathbf{1}\otimes\mathcal{E}\otimes\mathbf{1})|\psi\rangle_{ABC}\big]\nonumber\\
&=&E_a\big[|\psi\rangle_{ABC}\big]C[(\mathbf{1}\otimes\mathcal{E})|\phi^+\rangle\langle\phi^+|\big]\nonumber\\
&=&E_a\big(\rho_{AB})C[(\mathbf{1}\otimes\mathcal{E})|\phi^+\rangle\langle\phi^+|\big],
\end{eqnarray}
where $|\phi^+\rangle$ is the bipartite maximally entangled state,
$|\phi^+\rangle=1/\sqrt{2}(|00\rangle+|11\rangle)$.
\end{theorem}

{\em Proof:} Suppose the channel $\mathcal{E}$ is denoted by the
Kraus operators $\{K_j\}$ with $\sum_jK_j^\dagger K_j=\mathbf{1}$.
For the tripartite pure state $|\psi\rangle_{ABC}$, there must be a
set of optimal measurement $\{M_i\}$ for Charlie, such that the EOA
$E_a\big[|\psi\rangle_{ABC}\langle\psi|\big]
=\sum_ip_iC(|\psi_i\rangle)$ with the optimal pure-state
decomposition
$|\psi_i\rangle\langle\psi|=\mathrm{Tr}_C\big[(\mathbf{1}\otimes\mathbf{1}\otimes\otimes
M_i)
|\psi\rangle_{ABC}\langle\psi|(\mathbf{1}\otimes\mathbf{1}\otimes
M^{\dagger}_i)\big]/p_i$, and
$p_i=\mathrm{Tr}\big[(\mathbf{1}\otimes M_i)
|\psi\rangle_{ABC}\langle\psi|(\mathbf{1}\otimes
M^{\dagger}_i)\big]$. Considering the final state
$\rho'=\sum_j(\mathbf{1}\otimes K_j\otimes\mathbf{1})
|\psi\rangle_{ABC}\langle\psi|(\mathbf{1}\otimes
K_j^\dagger\otimes\mathbf{1})$, Charlie can also take the local
measurement $\{M_i\}$ and correspondingly obtains the states shared
by Alice and Bob,
$\rho'_i=\mathrm{Tr}_C\big[\sum_j(\mathbf{1}\otimes K_j\otimes M_i)
|\psi\rangle_{ABC}\langle\psi|(\mathbf{1}\otimes K_j^\dagger\otimes
M^{\dagger}_i)\big]/p'_i =\sum_j(\mathbf{1}\otimes
K_j)|\psi_i\rangle\langle\psi_i|(\mathbf{1}\otimes
K_j^\dagger)/p'_i$, and
$p'_i=\mathrm{Tr}\big[\sum_j(\mathbf{1}\otimes K_j\otimes M_i)
|\psi\rangle_{ABC}\langle\psi|(\mathbf{1}\otimes K_j^\dagger\otimes
M^{\dagger}_i)\big]=p_i$. Therefore, in virtue of the definition of
EOA for mixed states, the EOA satisfies \cite{expl}
\begin{eqnarray}
\label{prof EOA 1}
E_a(\rho')&\!\geq\!&\sum_ip_iC\big[(\mathbf{1}\otimes\mathcal{E})|\psi_i\rangle\big]\nonumber\\
&\!=\!&\sum_ip_iC(|\psi_i\rangle)C\big[(\mathbf{1}\otimes\mathcal{E})|\phi^+\rangle\langle\phi^+|\big]\nonumber\\
&\!=\!&E_a(|\psi\rangle_{ABC})C\big[(\mathbf{1}\otimes\mathcal{E})|\phi^+\rangle\langle\phi^+|\big],
\end{eqnarray}
where $|\phi^+\rangle$ is the 2-qubit maximally entangled state.

We next prove the reverse inequality also holds. We first assume that there is a set of optimal measurement
$\{M''_i\}$ for Charlie such that
\begin{equation}
\label{prof EOA 2} E_a(\rho')=\sum_ip''_i C(\rho''_i),
\end{equation}
where $p''_i=\mathrm{Tr}\big[\sum_j(\mathbf{1}\otimes K_j\otimes M''_i)
|\psi\rangle_{ABC}\langle\psi|(\mathbf{1}\otimes K_j^\dagger\otimes M''^{\dagger}_i)\big]$,
and
\begin{eqnarray}
&&\rho''_i\nonumber\\
&\!=\!&\frac{\mathrm{Tr}_C\sum_j(\mathbf{1}\otimes K_j\otimes M''_i)
|\psi\rangle_{ABC}\langle\psi|(\mathbf{1}\otimes K_j^\dagger\otimes M''^{\dagger}_i)}{p''_i}\nonumber\\
&\!=\!&\sum_j(\mathbf{1}\otimes K_j)
\tau''_i(\mathbf{1}\otimes K_j^\dagger).
\end{eqnarray}
Here $\tau''_i=\mathrm{Tr}_C((\mathbf{1}\otimes \mathbf{1}\otimes M''_i)
|\psi\rangle_{ABC}\langle\psi|(\mathbf{1}\otimes \mathbf{1}\otimes M''^{\dagger}_i))/p''_i$.
In terms of the results in Ref. \cite{thomas}, the inequality
$C((\mathbf{1}\otimes\mathcal{E})\rho)\leq C(\rho)C((\mathbf{1}\otimes\mathcal{E})|\phi^+\rangle)$
holds for a 2-qubit mixed state $\rho$. We thus have
\begin{eqnarray}
\label{prof EOA 3}
E_a(\rho')&\!=\!&\sum_ip''_i C(\rho''_i)\nonumber\\
&\!=\!&\sum_ip''_i C((\mathbf{1}\otimes\mathcal{E})\tau''_i)\nonumber\\
&\!\leq\!&C((\mathbf{1}\otimes\mathcal{E})|\phi^+\rangle\langle\phi^+|)\sum_ip''_i C(\tau''_i).
\end{eqnarray}
Due to
$\rho_{AB}=\mathrm{Tr}_C\big(|\psi\rangle_{ABC}\langle\psi|\big)
=\sum_ip''_i\tau''_i$ and the concavity of EOA for pure states, the
following equation holds,
\begin{eqnarray}
E_a(|\psi\rangle_{ABC})\geq\sum_ip''_iC(\tau''_i).
\end{eqnarray}
Accordingly, we have
\begin{eqnarray}
\label{prof EOA 4}
E_a(\rho')&\!\leq\!&C((\mathbf{1}\otimes\mathcal{E})|\phi^+\rangle\langle\phi^+|)\sum_ip''_i C(\tau''_i)\nonumber\\
&\!\leq\!&E_a(|\psi\rangle_{ABC})C((\mathbf{1}\otimes\mathcal{E})|\phi^+\rangle\langle\phi^+|).
\end{eqnarray}
Therefore, in terms of Eqs. (\ref{prof EOA 1}) and (\ref{prof EOA
4}), we obtain the equation,
\begin{eqnarray}
&&E_a\big[(\mathbf{1}\otimes\mathcal{E}\otimes\mathbf{1})|\psi\rangle_{ABC}\big]\nonumber\\
&\!=\!&E_a\big(|\psi\rangle_{ABC})C[(\mathbf{1}\otimes\mathcal{E})|\phi^+\rangle\langle\phi^+|\big]
\end{eqnarray}

The EOA reduction, with a one-sided noisy channel operating on Alice
or Bob subsystem, is independent of the initial state
$|\psi\rangle_{ABC}$ and completely determined by the channel¡¯s
action on the 2-qubit maximally entangled state. Thus, if we know
the dynamics of the Bell state¡¯s entanglement under a one-sided
noisy channel, we then know the time evolution of EOA for any pure
$2\times 2\times n_3$ initial states.

The result (\ref{theo1}) can also be generalized for a mixed $2\times 2\times n_3$ initial state $\rho_0$.
\begin{cor}
For a mixed $2\times 2\times n_3$ initial state $\rho_0$, with a
one-sided noisy channel operating on Bob subsystem,  we obtain its
evolution equation of EOA,
\begin{eqnarray}
\label{cor1}
&&E_a((\mathbf{1}\otimes\mathcal{E}\otimes\mathbf{1})\rho_0)\nonumber\\
&\!\leq\!&E_a(\rho_0)C[(\mathbf{1}\otimes\mathcal{E})|\phi^+\rangle\langle\phi^+|\big].
\end{eqnarray}
\end{cor}

This corollary is proved as follows. Suppose there exists a set of
optimal measurements for Charlie to create the state between Alice
and Bob, $\{q_i,\tau_i\}$, such that the EOA for the state $\rho_0$
satisfies the equation $E_a(\rho_0)=\sum_iq_iC(\tau_i)$. After the
channel operates on the initial state, we assume that there is
another set of optimal measurements for Charlie so that
$E_a((\mathbf{1}\otimes\mathcal{E}\otimes\mathbf{1})\rho_0)=\sum_iq'_iC((\mathbf{1}\otimes\mathcal{E})\tau'_i)$.
Due to $\sum_iq_i\tau_i=\mathrm{Tr}_C(\rho_0)
=\mathrm{Tr}_C[(\mathbf{1}\otimes\mathcal{E}\otimes\mathbf{1})\rho_0]=\sum_iq'_i\tau'_i$,
both $\{q_i,\tau_i\}$ and $\{q'_i,\tau'_i\}$ can be realized by the
Charlie's measurements no matter whether the local quantum channel
operates on the initial state. Therefore, we can prove
$E_a((\mathbf{1}\otimes\mathcal{E}\otimes\mathbf{1})\rho_0)=\sum_iq'_iC((\mathbf{1}\otimes\mathcal{E})\tau'_i)
\!\leq\!C[(\mathbf{1}\otimes\mathcal{E})|\phi^+\rangle\langle\phi^+|]\sum_iq'_iC(\tau'_i)
\!\leq\!C[(\mathbf{1}\otimes\mathcal{E})|\phi^+\rangle\langle\phi^+|]\sum_iq_iC(\tau_i)
\!=\!E_a(\rho_0)C[(\mathbf{1}\otimes\mathcal{E})|\phi^+\rangle\langle\phi^+|]$
by virtue of the concavity of EOA and the result,
$C[(\mathbf{1}\otimes\mathcal{E})\rho]\leq
C(\rho)C[(\mathbf{1}\otimes\mathcal{E})
|\phi^+\rangle\langle\phi^+|]$ with $\rho$ a 2-qubit state
\cite{thomas}.

This inequality (\ref{cor1}) holds for all one-sided channels $\mathcal{E}$. However,
assisted entanglement to prepare the entanglement between distant parties sometimes
needs to consider the case of two one-sided channels. In fact, the theorem 1
has an immediate generalization for two one-sided channels.
\begin{cor}
For a two one-sided channel $\mathcal{E}_1\times\mathcal{E}_2$
acting on Alice and Bob subsystem of a $2\times 2\times n_3$ state,
the evolution equation of EOA  leads to an inequality, which
provides an upper bound,
\begin{eqnarray}
\label{cor2}
&&E_a((\mathcal{E}_1\times \mathcal{E}_2\times\mathbf{1})\rho_0)\nonumber\\
&\!\leq\!&E_a(\rho_0)C[(\mathbf{1}\otimes\mathcal{E}_2)\!|\phi^+\rangle\!\langle\phi^+|]
C[(\mathcal{E}_1\otimes\mathbf{1})\!|\phi^+\rangle\!\langle\phi^+|].
\end{eqnarray}
\end{cor}

Due to $\mathcal{E}_1\times\mathcal{E}_2\times\mathbf{1}=(\mathbf{1}\otimes\mathcal{E}_2
\times\mathbf{1})(\mathcal{E}_1\times\mathbf{1}\times\mathbf{1})$,
we can prove this corollary easily by virtue of the corollary 2.

Although (\ref{cor1}) and (\ref{cor2}) are not equalities, the
evolution equation of EOA (\ref{theo1}) and these upper bounds
(\ref{cor1},\ref{cor2}) are only concerned with the quantity, the
entanglement of the final state evolved from a 2-qubit maximally
entangled state with the channel's action on it. Hence, these
results will ease the experimental characterization of entanglement
dynamics of EOA under unknown channels in a preparation of 2-qubit
state by assisted entanglement. We, instead of exploring the
time-dependent action of the channel on all initial states, only
need to investigate the entanglement evolution of the maximally
entangled state alone.

The assisted entanglement can also produce a general bipartite
entangled state besides 2-qubit states. Moreover, bipartite states
with higher dimension can improve the performance of various quantum
information and computation tasks. Hence, it is necessary to explore
the dynamics of EOA for the characterization of the produced
entanglement produced by assisted entanglement when the state is coupled to its environment.
In the following section, we
explore how EOA evolves for a $d\! \times  d \! \times n_3$ state with
local noisy channels operation on it.

\section{\bf Evolution equation of EOA for $d\times d\times n_3$ systems}
First, we need to define EOA for a $d\times d\times n_3$ state.
For the consistency with the definition of EOA, we define
entanglement of assistance in terms of the entanglement measure
I-concurrence \cite{rungta,li2}:
\begin{eqnarray}
\label{EOA2}
E_a(|\psi\rangle_{ABC})
&\!\equiv\!&\textrm{max}\sum_ip_iC(|\phi_i\rangle_{AB})\nonumber\\
&\!=\!&\textrm{max}\sum_ip_i \sqrt{\sum_{mn}|\langle\phi_i^*|
S_{mn}|\phi_i\rangle|^2},
\end{eqnarray}
which is maximized over all possible pure-state decompositions of
$\rho_{AB}=\textrm{Tr}_C[|\psi\rangle_{ABC}\langle\psi|]=\sum_ip_i|\phi_i\rangle_{AB}\langle\phi_i|$.
where $S_{mn}=L_m\otimes L_n$, $L_m, m=1,..., d(d-1)/2$, $L_n, n=1,
..., d(d-1)/2$ are the generators of group $SO(d)$. In the same way,
EOA, for a mixed tripartite state $\rho_{ABC}$, is defined as the
maximum achievable average I-concurrence shared by Alice and Bob
after Charlie has performed a local operation and communicated the
result to Alice and Bob.

For a general $\rho_{ABC}$ with Bob's system undergoing the action of a noisy channel
$\mathcal{E}$, the final state takes the form
$\rho'_{ABC}=(\mathbf{1}\otimes\mathcal{E}\otimes\mathbf{1})\rho_{ABC}$. Correspondingly,
the evolution equation of EOA is obtained as the following theorem.
\begin{theorem}
The EOA of the final state $\rho'_{ABC}$ satisfies the equation,
\begin{eqnarray}
\label{theo2}
&&E_a\big[(\mathbf{1}\otimes\mathcal{E}\otimes\mathbf{1})\rho_{ABC}\big]\nonumber\\
&\!\leq\!&\frac{d}{2}E_a(\rho_{ABC})C[(\mathbf{1}\otimes\mathcal{E})|\chi^+\rangle\langle\chi^+|\big]
\end{eqnarray}
where $|\chi^+\rangle$ is a $d\times d$ maximally entangled state,
$|\chi^+\rangle=\!\sum_{i=0}^{d-1}|i\rangle\otimes|i\rangle/\sqrt{d}$.
\end{theorem}

{\em Proof:} In order to prove (\ref{theo2}), we need the result in
Ref. \cite{li}. For a $d\times d$ bipartite state $\rho$ with its
second subsystem going through a noisy channel $\mathcal{E}$, we
have $C\big[(\mathbf{1}\otimes\mathcal{E})\rho\big]\leq
\frac{d}{2}C(\rho)C\big[(\mathbf{1}\otimes\mathcal{E})|\chi^+\rangle\langle\chi^+|\big].$

We first assume that there is a set of optimal measurement
$\{M_i\}$ for Charlie such that
\begin{eqnarray}
\label{theo2-1}
E_a[(\mathbf{1}\otimes\mathcal{E}\otimes\mathbf{1})\rho_{ABC}\big]=\sum_ip_i
C(\rho_i),
\end{eqnarray}
where $p_i=\mathrm{Tr}\big[\sum_j(\mathbf{1}\otimes K_j\otimes M_i)
\rho_{ABC}(\mathbf{1}\otimes K_j^\dagger\otimes M^{\dagger}_i)\big]$,
and
\begin{eqnarray}
\label{theo2-2}
\rho_i
&\!=\!&\frac{\mathrm{Tr}_C\sum_j(\mathbf{1}\otimes K_j\otimes M_i)
\rho_{ABC}(\mathbf{1}\otimes K_j^\dagger\otimes M^{\dagger}_i)}{p_i}\nonumber\\
&\!=\!&\sum_j(\mathbf{1}\otimes K_j)
\tau_i(\mathbf{1}\otimes K_j^\dagger).
\end{eqnarray}
Here $\tau_i=\mathrm{Tr}_C((\mathbf{1}\otimes \mathbf{1}\otimes M_i)
\rho_{ABC}(\mathbf{1}\otimes \mathbf{1}\otimes M^{\dagger}_i))/p_i$.
Therefore, we have
\begin{eqnarray}
\label{theo2-3}
&&E_a\big[(\mathbf{1}\otimes\mathcal{E}\otimes\mathbf{1})\rho_{ABC}\big]\nonumber\\
&\!=\!&\sum_ip_iC((\mathbf{1}\otimes\mathcal{E})\tau_i)\nonumber\\
&\!\leq\!&\frac{d}{2}C[(\mathbf{1}\otimes\mathcal{E})|\chi^+\rangle\langle\chi^+|\big]\sum_ip_iC(\tau_i) \nonumber\\
&\!\leq\!&\frac{d}{2}C[(\mathbf{1}\otimes\mathcal{E})|\chi^+\rangle\langle\chi^+|\big]E_a(\rho_{ABC}).
\end{eqnarray}
Here the last inequality is obtained by the concavity of EOA. From the above analysis, we
have proved the theorem.

{\em Remark.} In particular, if the initial state is a $d\times
2\times n_3$ pure state, we will obtain an evolution equation of EOA
for such a state, $
E_a((\mathbf{1}\otimes\mathcal{E}\otimes\mathbf{1})|\phi\rangle_{ABC}\langle\phi|)
\!=\!E_a(|\phi\rangle_{ABC})C[(\mathbf{1}\otimes\mathcal{E})|\phi^+\rangle\langle\phi^+|\big]
$. By the aid of the following equation \cite{li},
$C[(\mathbf{1}\otimes\mathcal{E})|\psi\rangle_{AB}\langle\psi|]
=C(|\psi\rangle_{AB}){C}[(\mathbf{1}\otimes\mathcal{E})|\phi^+\rangle\langle\phi^+|]$
with $|\psi\rangle_{AB}$ a pure $n_1\times 2$ state, the evolution
equation is then proved in a way similar to that of Theorem 1. If we
consider a one-sided channel $\mathcal{E}$ which operates on the
third subsystem, we get an lower bound for the EOA of the final
state, $E_a\big[(\mathbf{1}\otimes\mathbf{1}\otimes\mathcal{E})\rho_{ABC}\big]\geq
C(\rho_{AB})$, where $\rho_{AB}=\mathrm{Tr}_C\rho_{ABC}$. We can
also generalize the Theorem 2 to the case of two one-sided channel
$\mathcal{E}_1\otimes\mathcal{E}_2$, whose result is similar to
(\ref{theo2}).

Similarly, the dynamics of EOA for a general
$d\times d\times n_3$ state, which goes through a local one-sided
noisy channel, is determined by the channel's action on the
bipartite maximally entangled state
$|\chi^+\rangle\!=\!\sum_{i=0}^{d-1}|i\rangle\otimes|i\rangle/\sqrt{d}$.
Therefore, it allows one to understand, without resorting to the
channel's action on all initial states, the dynamics of EOA under
some local quantum channels for an arbitrary initial state, if one
knows the time evolution of entanglement for the bipartite maximally
entangled state under such quantum channels.

\section{\bf EXAMPLES AND DISCUSSION}
Let us consider several examples. Suppose $\mathcal{E}$ is a phase
damping channel, such that $\mathcal{E}(\rho)=M_0\rho
M_0^\dag+M_1\rho M_1^\dag$ with $M_0=\tiny{\left(
   \begin{array}{cc}
   1& 0 \\
   0 &\nu
   \end{array}\right)},
   M_1=\tiny{\left(
   \begin{array}{cc}
   0& 0 \\
   0 &\omega
   \end{array}\right)}$, where the time-dependent parameters
$\nu\!=\!\mathrm{exp}[-\Gamma t]$ and $\omega\!=\!\sqrt{1-\nu^2}$.
For a generalized GHZ state
$|\psi_g\rangle_{ABC}\!=\!\alpha|000\rangle\!+\!\sqrt{1-|\alpha|^2}|111\rangle$,
we can get
$E_a\big[(\mathbf{1}\otimes\mathcal{E}\otimes\mathbf{1})|\psi_g\rangle_{ABC}\langle\psi_g|\big]
\!=\!C[(\mathbf{1}\otimes\mathcal{E})|\phi^+\rangle\langle\phi^+|]E_a(|\psi_g\rangle_{ABC})
=2\mathrm{exp}[-\Gamma t]\alpha \sqrt{1-|\alpha^2|}$ in terms of Eq.
(\ref{theo1}). Next we suppose the quantum channel is a generalized
amplitude damping channel, $\mathcal{E}_{GAD}$, describing the
effect of dissipation to an environment at finite temperature. The
channel usually adopts the form as: $K_0=\sqrt{p}\tiny{\left(
   \begin{array}{cc}
   1& 0 \\
   0 &\nu
   \end{array}\right)},
K_1=\sqrt{p}\tiny{\left(
   \begin{array}{cc}
   0& \omega \\
   0 & 0
   \end{array}
\right )},
K_2=\sqrt{1-p}\tiny{\left(
   \begin{array}{cc}
   \nu& 0 \\
   0 & 1
   \end{array}
\right )},
K_3=\sqrt{1-p}\tiny{\left(
   \begin{array}{cc}
   0& 0 \\
   \omega & 0
   \end{array}
\right)} $. Without loss of generality, we set $p=\frac{1}{2}$, then
obtain
$E_a\big[(\mathbf{1}\otimes\mathcal{E}_{GAD}\otimes\mathbf{1})|\psi_g\rangle_{ABC}\langle\psi_g|\big]
=C^2((\mathcal{E}_{GAD}\otimes\mathbf{1})|\phi^+\rangle\langle\phi^+|)E_a(|\psi_g\rangle_{ABC})
=|\alpha|\sqrt{1-|\alpha|^2} (\mathrm{exp}[-2\Gamma
t]+2\mathrm{exp}[-\Gamma t]-1)$.
\begin{figure}[!b]
\begin{center}
\scalebox{0.56}[0.5]{\includegraphics*[0pt,0pt][370pt,210pt]{eoa1.eps}}
\caption{The decay of
$E_a\big[(\mathbf{1}\otimes\mathcal{E}\otimes\mathbf{1})|\psi_g\rangle_{ABC}\langle\psi_g|\big]$
vs $\Gamma t$, where $\Gamma$ is a generalized amplitude decay rate.
The dashed line and solid line are the decay of EOA for phase
damping channel and generalized amplitude damping channel
respectively with $\alpha=\frac{1}{2}$. Note that in solid line a
sudden death of EOA appears but in dashed line it does not.}
\end{center}
\end{figure}
From the above equation, we can find there is a sudden death of EOA
for this damping channel, similar to the sudden death of
entanglement \cite{yuting1,mp,yuting2}. Just like the sudden death
of entanglement cannot appear for any channel \cite{yuting1}, the
sudden death of EOA does not exist for some channels. As shown in
Fig. 1, the dashed line, the evolution of
$E_a\big[(\mathbf{1}\otimes\mathcal{E}\otimes\mathbf{1})|\psi_g\rangle_{ABC}\langle\psi_g|\big]$
with $\mathcal{E}$ a phase damping channel,
indicates $E_a$ does not die suddenly but asymptotically. However,
for the generalized amplitude damping channel, the EOA does go
abruptly to zero in a finite time and remain zero thereafter. From
the above analysis, we can find, under the local quantum channel's
action, whether the sudden death of EOA appears depends only on the
channel's action on maximally entangled states.

The inequality
$C^2_a(\rho_{AB})+C^2_a(\rho_{AC})-C^2(|\psi\rangle_{A\!-\!BC})
=C^2(|\psi\rangle_{A\!-\!BC})-C^2(\rho_{AB})-C^2(\rho_{AC})\geq0$
holds \cite{gour1}. First, we denote the quantity by
$\tau(\rho_{ABC})$,
$\tau(\rho_{ABC})=C^2_a(\rho(AB))+C^2_a(\rho(AC))-C^2(\rho_{A\!-\!BC})$,
which is just the three-tangle \cite{coffman} if $\rho_{ABC}$ is
pure. We consider a case that a mixed state $\rho$ can be rewritten
as
$\rho=\sum_(K_i\otimes\mathbf{1}\otimes\mathbf{1})|\psi\rangle_{ABC}
\langle\psi|(K^\dagger_i\otimes\mathbf{1}\otimes\mathbf{1})$ with
$\{K_i\}$ representing the operation of a quantum channel
$\mathcal{E}$, i.e. an output state evolved from the initial state
$|\psi\rangle_{ABC}$ of which the first subsystem goes through a
quantum channel $\mathcal{E}$. Then, according to the Theorem 1 and
the result in Ref. \cite{li}, we obtain the following equations,
$\tau((\mathcal{E}\otimes\mathbf{1}\otimes\mathbf{1})|\psi\rangle_{ABC}\langle\psi|)
=C^2_a(\rho(AB))+C^2_a(\rho(AC))-C^2(\rho_{A\!-\!BC})
=C^2((\mathcal{E}\otimes\mathbf{1})|\phi^+\rangle\langle\phi^+|)[C^2_a(\rho_{AB})+C^2_a(\rho_{AC})-C^2(|\psi\rangle_{A\!-\!BC})]
=C^2((\mathcal{E}\otimes\mathbf{1})|\phi^+\rangle\langle\phi^+|)\tau(|\psi\rangle_{ABC})\geq0$.
Up to now, however, no references provide a proof whether or not the
inequality
$C^2_a(\rho(AB))+C^2_a(\rho(AC))-C^2(\rho_{A\!-\!BC})\geq0$ is
always satisfied for a general mixed 3-qubit state $\rho_{ABC}$.
Although we only prove the quantity $\tau$ is greater than or equal
to zero for such a mixed 3-qubit state, we are willing to conjecture
that it is still valid for any mixed 3-qubit states.

\section{\bf Summary}
In summary, we have investigated the time evolution of EOA when one
subsystem undergoes the action of an arbitrary noisy channel. In
particular, for a $2\times 2\times n_3$ pure quantum state with the
qubit subsystem being subject to a noisy channel, we get a general
factorization law for evolution equation of EOA. Furthermore, for
the other cases with one subsystem undergoing the action of local
quantum channels, a similar relation is satisfied, at a price of
turning the factorization law into an inequality.  We use
generalized amplitude damping and phase damping channels as examples
and find that the sudden death of EOA does exist in the evolution
and is only determined by the entanglement evolution equation of the
maximally entangled state entering this channel. As a result, in
order to characterize the dynamics of EOA, we, instead of exploring
the time-dependent action of the channel on all initial states, only
need to investigate the entanglement evolution of the maximally
entangled state, which is universal for all initial states passing
through this channel. Hence, these results will ease the
experimental characterization of entanglement dynamics of EOA under
unknown channels in an experimental preparation of a bipartite state
by assisted entanglement.

This work was supported by NSFC under Grants No. 10874235,
10934010, and 60978019, the NKBRSFC under Grants No. 2006CB921400,
2009CB930704, 2010CB922904 and KZ200810028013.

\end{document}